# A Load Impedance Emulation Active Interface for Piezoelectric Vibration Energy Harvesters

**Alessandro Lo Schiavo, Luigi Costanzo and Massimo Vitelli**

Department of Engineering, Università degli Studi della Campania Luigi Vanvitelli, Aversa, CE, Italy
E-mail: alessandro.loschiavo@unicampania.it, luigi.costanzo@unicampania.it, and massimo.vitelli@unicampania.it

**Abstract**

A single stage active AC/DC interface able to emulate the optimal load impedance of a Resonant Piezoelectric Vibration Energy Harvester (RPVEH) is proposed. As theoretically shown, unlike an electronic interface that emulates an optimal load generator, an interface that emulates an optimal load impedance does not require adaptation to the acceleration of input vibrations. This allows the use of a very simple control, avoiding the implementation of Maximum Power Point Tracking (MPPT) algorithms that require lossy microcontrollers. Thus, the proposed interface is equipped with a simple analog controller allowing the RPVEH to work in its Maximum Power Point (MPP) in both steady-state and variable conditions of vibrations, without recurring to multivariable perturbative approaches, as it happens for the most of single stage AC/DC interfaces proposed in the literature. The absence of perturbative techniques allows a significant improvement of both stationary and dynamic performances. Experimental tests of a prototype of the proposed interface confirm the theoretical findings and the predicted behavior.

## 1 Introduction

The most widespread AC/DC interface for vibration energy harvesters, both in laboratory prototypes [1]-[3] and in commercial devices [4]-[7], is the diode bridge rectifier due to the advantage of being very simple and suitable for low power applications. However, the performance of a passive diode bridge rectifier depends on the harvester characteristics and often these characteristics do not allow a passive rectifier to extract the maximum available power [1]. Therefore, highly performing devices and techniques have been proposed to overcome the limitations of passive rectifiers and increase the extractable power. Noteworthy examples are the interfaces based on the Synchronized Switching Harvesting on Inductor (SSHI) [8]-[9], the Synchronous Electric Charge Extraction (SECE) [10]-[11], or the Energy Harvester Power Optimizer (EHPO) [12]. In these cases, the AC/DC power electronic interfaces are composed of multiple stages that impact the overall system efficiency [1]. Moreover, in some cases such interfaces require single variable Maximum Power Point Tracking (MPPT) techniques to extract the maximum available power and to efficiently operate in case of variable input vibrations.

Alternative solutions are represented by single stage active AC/DC switching converters that are able to extract the maximum available power whatever is the energy harvester [13]-[19]. However, a disadvantage of these interfaces is represented by the complexity of their control that, in many cases, is implemented by means of a microcontroller, with a negative impact on the overall energetic performance. Indeed, they typically need multi variable MPPT techniques, which are based on complex perturbative approaches, in order to identify and track the Maximum Power Point (MPP). Such MPPT techniques are characterized by steady-state oscillations around the MPP and by reduced dynamic performance compared to single variable MPPT techniques. All these aspects impact the energetic performance of the overall system.

In this paper a single stage active AC/DC interface able to emulate the optimal load impedance of a Resonant Piezoelectric Vibration Energy Harvester (RPVEH) is proposed, with the aim of overcoming the above limitations thanks to two main features. Firstly, as it will be theoretically investigated and experimentally shown, the emulation of an optimal load impedance for the RPVEH, instead of an optimal load generator, allows the use of a simple analog controller, without resorting to a microcontroller with its energy losses.

Secondly, the emulation of an optimal load impedance does not require an adaptation to the vibration conditions, but it allows the RPVEH to work in its MPP in both stationary and dynamic conditions. The lack of adaptation, usual in the perturbation approaches of MPPT techniques, avoids the oscillations around the MPP in steady state conditions improving the stationary performance. Moreover, the lack of adaptation ensures high dynamic performance in presence of input acceleration variations.

The rest of the paper is organized as it follows. In Section 2, a preliminary analysis of the optimal loads for piezoelectric harvesters and of the maximum extractable power is reported. In Section 3, the differences between a load impedance and a load generator are investigated to assess the strengths and the weaknesses in harvesting applications. In Section 4, the proposed active interface is presented and in Section 5, a prototype of the proposed interface is experimentally tested. Conclusions end the paper.

## 2 Optimal Loads for Piezoelectric Harvesters

Let us consider a Resonant Piezoelectric Vibration Energy Harvester (RPVEH) in a cantilever configuration as represented in Fig. 1(a). According to the widely used linear single degree of freedom lumped model, the RPEVH can be divided in a mechanical stage and an electrical stage as shown in Fig. 1(b) and it can be schematized by using the equivalent electric circuit shown in Fig. 1(c) [1]. The spring stiffness is denoted by $K$, the vibrating mass is denoted by $M$, the viscous damping coefficient by $D$ and the force factor describing the piezoelectric effect by $\alpha$ (it has units [N/V]). The mechanical stage is represented by a spring-mass-damper resonant system corresponding in the equivalent resonant circuit to $C_K = \alpha^2/K$, $L_M = M/\alpha^2$ and $R_D = D/\alpha^2$. The input vibration acceleration is $\ddot{y}(t)$, and $\Delta = M/\alpha$ (it has units [V/g]). With reference to the electrical stage, it is made of a generator imposing the current $i_{piezo}(t)$ placed in parallel with the equivalent output capacitance $C_p$ of the piezoelectric layers. As shown in Fig. 1, $i_{piezo}(t) = \alpha \cdot \dot{x}(t)$ is the current generated by the piezoelectric effect that is a function of the speed of the cantilever tip $\dot{x}(t)$. The RPVEH mechanical resonance angular frequency is $\omega_{res} = \sqrt{K/M}$. Usually cantilever RPVEHs have very narrow 3dB bandwidths around $\omega_{res}$, hence, the following analysis will be focused on the harvester behaviour around the resonance frequency.

Let's consider a vibration acceleration $\ddot{y}(t) = A_{max} \cdot sin(\omega_{res} \cdot t)$, with amplitude $A_{max}$ and angular frequency $\omega_{res}$.

In the following analysis, the quality factor $Q$ of the mechanical resonator, and the coupling coefficient $\rho$ between the electrical capacitance $C_p$ and the equivalent mechanical

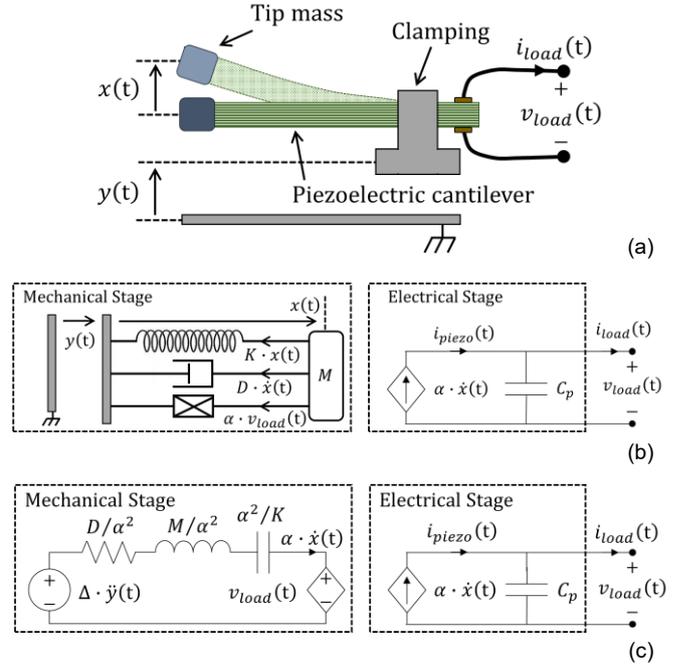

Fig. 1. (a) Resonant Piezoeletric Vibration Energy Harvester (RPVEH) in a cantilever configuration. (b) Spring-mass-damper model of the mechanical stage and equivalent circuit of the electrical stage. (c) Equivalent electric circuit of a RPVEH

capacitance $\alpha^2/K$, defined as it follows, will be considered

$$Q = \frac{K}{\omega_{res} \cdot D}; \qquad \rho = \frac{\omega_{res} \cdot D \cdot C_p}{\alpha^2} = \frac{1}{Q}\frac{K \cdot C_p}{\alpha^2} \qquad (1)$$

The equivalent output impedance of the harvester can be expressed as $\dot{Z}_p = R_P + jX_P$, where

$$R_p = \frac{1}{\omega_{res}C_p} \cdot \frac{\rho}{1+\rho^2}; \qquad X_p = -\frac{1}{\omega_{res}C_p} \cdot \frac{\rho^2}{1+\rho^2} \qquad (2)$$

The harvester open circuit voltage has angular frequency $\omega_{res}$, amplitude $V_{oc}$ and phase $\Phi_{oc}$, that is $v_{oc}(t) = V_{oc} \cdot sin(\omega_{res} \cdot t + \Phi_{oc})$. Its representation in the phasor domain is the following

$$\bar{V}_{oc} = V_{oc} \cdot e^{j\Phi_{oc}} = \frac{\Delta \cdot A_{max}}{1 + j \cdot \rho} \qquad (3)$$

Since an electronic interface connected to the harvester terminals can emulate either an impedance or a voltage generator, it is worth calculating the power extracted by the harvester in these two cases, as shown in Fig. 2.

Let's focus initially on the load impedance emulation case, and in particular, let's consider a parallel connection between a resistance $R_{load}$ and a reactance $X_{load}$, as shown in Fig. 2(a).

The average extracted power $P_{Z-load}$, which is a function of $A_{max}$, $R_{load}$ and $X_{load}$, is

$$P_{Z-load}(A_{max}, R_{load}, X_{load}) = \frac{(\Delta \cdot A_{max})^2}{2} \frac{\omega_{res} C_p}{\rho} \frac{\Psi_Z}{(1+\Psi_Z)^2 + \left(\Psi_Z \frac{R_{load}}{X_{load}} - \rho\right)^2} \quad (4)$$

where

$$\Psi_Z(R_{load}, X_{load}) = \frac{\omega_{res} C_p (1+\rho^2)}{\rho} \frac{R_{load} X_{load}^2}{R_{load}^2 + X_{load}^2} \quad (5)$$

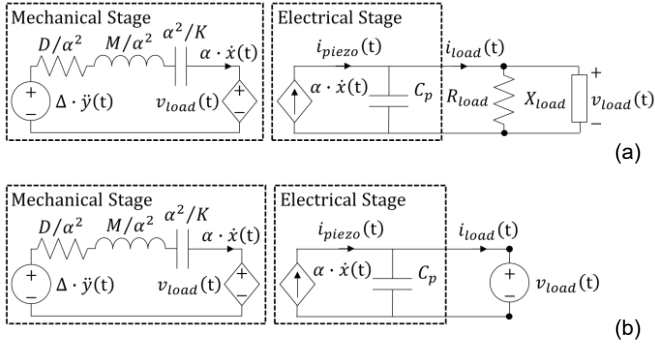

Fig. 2. (a) RPVEH connected to a load impedance in a parallel connection. (b) RPVEH connected to a load generator.

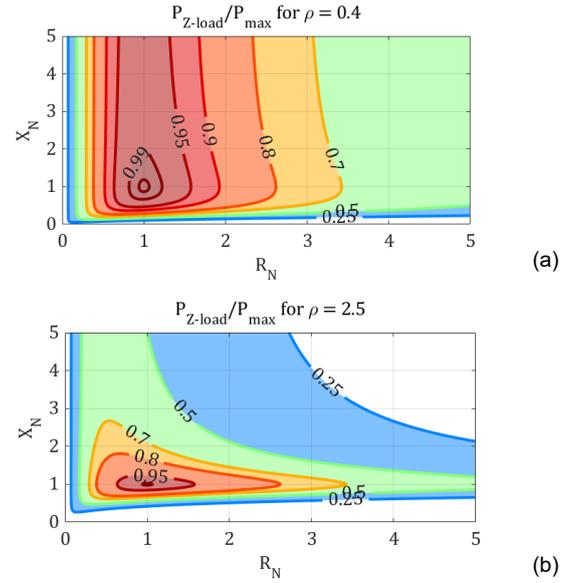

Fig. 3. Contour plots of the normalized power $P_{Z-load}/P_{max}$ as a function of $R_N$ and $X_N$ for two values of $\rho$. (a) $\rho = 0.4$; (b) $\rho = 2.5$.

The maximum power transfer theorem provides the conditions for the maximization of the average extracted power, which is a fundamental target in the design of energy harvesting systems. According to such a theorem, the optimal load impedance that maximizes the extracted power is

$$\dot{Z}_{opt} = \dot{Z}_p^* = \frac{1}{\omega_{res} C_p} \cdot \frac{\rho}{1+\rho^2}(1+j\cdot\rho) \quad (6)$$

where the asterisk (*) denotes the complex conjugate. When connected to the optimal impedance $\dot{Z}_{opt}$, the RPVEH output voltage and the maximum extracted power take the following expressions

$$\bar{V}_{opt} = \bar{V}_{oc} \cdot \frac{\dot{Z}_{opt}}{\dot{Z}_p + \dot{Z}_{opt}} = \frac{\Delta \cdot A_{max}}{2} \quad (7)$$

and

$$P_{max} = \frac{(\Delta \cdot A_{max})^2}{8} \frac{\omega_{res} C_p}{\rho} = \frac{(\Delta \cdot A_{max})^2}{8} \frac{\alpha^2}{D} \quad (8)$$

By taking into account (8), it is possible to express (4) as a function of the maximum extractable power $P_{max}$

$$P_{Z-load}(A_{max}, R_N, X_N) = P_{max} \cdot \frac{4 \cdot \Psi_N X_N}{(1+\Psi_N X_N)^2 + \rho^2(\Psi_N R_N - 1)^2} \quad (9)$$

where

$$\Psi_N(R_N, X_N) = \frac{(1+\rho^2) R_N X_N}{(R_N^2 \cdot \rho^2 + X_N^2)} \quad (10)$$

$R_N = R_{load}/R_{opt}$ and $X_N = X_{load}/X_{opt}$, being $R_{opt}$ and $X_{opt}$ the values of $R_{load}$ and of $X_{load}$ that maximize the extracted power in (9), i.e.

$$R_{opt} = \frac{\rho}{\omega_{res} C_p} = \frac{D}{\alpha^2} \quad (11.1)$$

$$X_{opt} = \frac{1}{\omega_{res} C_p} \quad (11.2)$$

It is possible to observe that, in addition to the normalized load resistance $R_N$ and the normalized load reactance $X_N$, the normalized power $P_{Z-load}/P_{max}$ depends only on $\rho$.

As an example, in Fig. 3, the contour plots of the normalized powers $P_{Z-load}/P_{max}$ are shown as a function of $R_N$ and $X_N$ for two values of $\rho$.

Let's now focus on the case of a sinusoidal voltage generator connected to the harvester terminals, as shown in Fig. 2(b). If the generator has amplitude $V_{load}$ and phase $\Phi_{load}$, its representation in the phasor domain is $\bar{V}_{load} = V_{load} \cdot e^{j\Phi_{load}}$. The phasor of the current flowing through the generator can be expressed as

$$\bar{I}_{load} = \frac{\bar{V}_{oc} - \bar{V}_{load}}{\dot{Z}_p} = \frac{\left(\frac{\Delta \cdot A_{max}}{1+j\cdot\rho} - \bar{V}_{load}\right)}{\frac{1}{\omega_{res} C_p} \cdot \frac{\rho}{1+\rho^2}(1-j\cdot\rho)} \quad (12)$$

In this case, the average extracted power, which is a function of $A_{max}$, $V_{load}$ and $\Phi_{load}$, is given by

$$P_{V-load}(A_{max}, V_{load}, \Phi_{load}) = \frac{(\Delta \cdot A_{max})^2}{2} \cdot \frac{\omega_{res} C_p}{\rho} \cdot \Psi_V \left[1 - \frac{\Psi_V}{cos^2(\Phi_{load})}\right] \quad (13)$$

where

$$\Psi_V(A_{max}, V_{load}, \Phi_{load}) = \frac{V_{load}}{\Delta \cdot A_{max}} \cdot cos(\Phi_{load}) \quad (14)$$

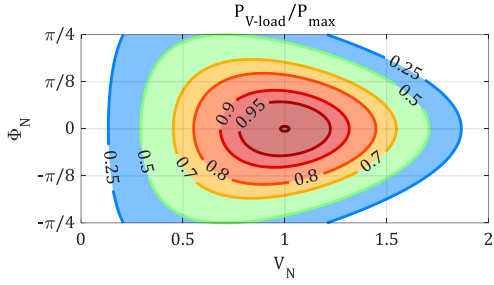

Fig. 4. Contour plot of the normalized power $P_{V-load}/P_{max}$ as a function of $V_N$ and $\Phi_N$

By taking into account (8), it is possible to express (13) as a function of the maximum extractable power $P_{max}$

$$P_{V-load}(A_{max}, V_N, \Phi_N) = $$
$$= P_{max} \cdot 2 \cdot V_N \cdot cos(\Phi_N)\left[1 - \frac{V_N}{2 \cdot cos(\Phi_N)}\right] \quad (15)$$

where $V_N = V_{load}/V_{opt}$ and $\Phi_N = \Phi_{load} - \Phi_{opt}$, being $V_{opt}$ and $\Phi_{opt}$ the values of $V_{load}$ and of $\Phi_{load}$ that maximize (15)

$$V_{opt} = \frac{\Delta \cdot A_{max}}{2} \quad (16.1)$$
$$\Phi_{opt} = 0 \quad (16.2)$$

It is very interesting to observe that, in the voltage generator emulation case, the normalized power $P_{V-load}/P_{max}$ does not depend on the RPVEH parameters but only depends on the normalized load voltage generator amplitude $V_N$ and on the normalized load voltage generator phase $\Phi_N$. In Fig. 4, the contour plot of $P_{V-load}/P_{max}$ is shown as a function of $V_N$ and $\Phi_N$.

It is worth highlighting that according to (15), the maximum power extracted by a harvester connected to an optimal voltage generator (that is (15) evaluated for $V_N = 1$ and $\Phi_N = 0$) is just equal to $P_{max}$, which is the same power extracted by the harvester when it is connected to an optimal impedance (that is (9) evaluated for $R_N = 1$ and $X_N = 1$). In other words, both an emulated load impedance of value (11) and an emulated load voltage generator of value (16) ensure the extraction of the maximum power, and they can be both considered as optimal loads.

## 3 Load Impedance versus Load Generator

Since an electronic interface connected to the harvester terminals can emulate both a load impedance and a load generator, it is worth investigating whether it is preferable for the electronic interface to behave like the impedance in Fig. 2(a) or like the voltage generator in Fig. 2(b). While in previous section it was shown that in both cases the same maximum power (8) can be extracted, in this section the differences between the two cases are highlighted.

The main difference, which has a significant practical impact, consists in the dependence of the optimal condition on the value of the acceleration amplitude. In the case of load generator, expressions (16) show that the optimal voltage amplitude $V_{opt}$ is a function of the input vibration acceleration, while in the case of load impedance, expressions (11) show that the optimal values do not depend on the acceleration amplitude.

Since the exact value of the acceleration amplitude $A_{max}$ is usually not known in the design stage and, in any case, it usually varies over time during the harvester operation, according to (16), also the optimal voltage changes over time. Therefore, if an electronic interface emulates a voltage generator, in order to extract the maximum available power, a dynamic MPPT control should be implemented to identify in real time the optimal voltage and to track the maximum power condition. Obviously, such a control involves additional power losses in the power electronic interface.

On the other hand, expressions (11) show that in case of load impedance emulation, the values of $R_{opt}$ and $X_{opt}$ are not a function of $A_{max}$ and thus they do not change over time. Therefore, once the optimal load impedance, which depends on the harvester characteristics only, has been identified, whatever is the acceleration amplitude, it is possible to always extract the maximum average power from the RPVEH without any modification.

Thus, in case of a load impedance, a dynamic MPPT control aimed at tracking the variations of the maximum power point is not needed. It is enough an initial identification of the optimal conditions given by (11) and dependent exclusively on the harvester characteristics.

Obviously, in order not to withstand the losses associated with the MPPT control in the case of load generator, the control could be omitted, and a voltage generator with constant amplitude and phase could be used, at the cost of extracting less power than the maximum when $A_{max}$ is different from the nominal one. In this scenario, it is interesting to quantify the impact of the absence of such an MPPT control on the extracted power, when the acceleration amplitude varies over time. To this aim, let's consider an RPVEH working in presence of a starting acceleration amplitude $A_{max-0}$ in correspondence of which it is possible to extract the maximum power given by (8) and equal to

$$P_{max-0} = \frac{(\Delta \cdot A_{max-0})^2}{8}\frac{\alpha^2}{D} \quad (17)$$

Such a power can be extracted both for an emulated load impedance given by (11) and for an emulated load generator given by (16), i.e.

$$V_{opt-0} = \frac{\Delta \cdot A_{max-0}}{2} \quad (18.1)$$
$$\Phi_{opt-0} = 0 \quad (18.2)$$

Let's now assume that the acceleration amplitude changes and becomes equal to $A_{max}$. In the case of the load impedance, without any modification of the load characteristics, the extracted power becomes

$$P_{load-Z0} = \frac{(\Delta \cdot A_{max})^2}{8} \frac{\alpha^2}{D} = P_{max}(A_{max}) \quad (19)$$

which is still the maximum extractable power.

In the case of the load generator, without any modification, the extracted power, given by (15), becomes

$$P_{load-V0} = P_{load}(A_{max}, V_{opt-0}, \Phi_{opt-0}) =$$
$$= P_{max} \cdot \frac{2 \cdot A_{max-0}}{A_{max}} \left[1 - \frac{A_{max-0}}{2 \cdot A_{max}}\right] \quad (20)$$

By substituting (19) in (20), it results

$$P_{load-V0} = P_{load-Z0} \cdot \frac{2 \cdot A_{max-0}}{A_{max}} \left[1 - \frac{A_{max-0}}{2 \cdot A_{max}}\right] \quad (21)$$

On the basis of (21), it is possible to define the following percentage coefficient $\lambda_{waste}$

$$\lambda_{waste} = \frac{P_{load-Z0} - P_{load-V0}}{P_{load-Z0}} =$$
$$= \left[1 - \frac{2 \cdot A_{max-0}}{A_{max}} \left(1 - \frac{A_{max-0}}{2 \cdot A_{max}}\right)\right] \cdot 100\% \quad (22)$$

which is always lower than 100%.

Equation (22) provides a quantitative estimation of the power reduction that, in presence of a variation of the vibration acceleration amplitude from $A_{max-0}$ to $A_{max}$, is obtained in the load generator case compared to the load impedance case, if no MPPT control is carried out.

Since such power reduction depends only on the acceleration amplitude, the trend of the powers $P_{max}$, $P_{load-Z0}$, and $P_{load-V0}$ (extracted in correspondence of $A_{max}$) normalized to the starting maximum power $P_{max-0}$ (extracted in correspondence of $A_{max-0}$) are reported in Fig. 5(a) as a function of the normalized acceleration amplitude $A_{max}/A_{max-0}$. In Fig. 5(b), the trend of $\lambda_{waste}$ is reported as a function of $A_{max}/A_{max-0}$. It is interesting to observe, as an example, that if the acceleration amplitude doubles ($A_{max}/A_{max-0} = 2$) the extracted power $P_{load-V0}$ with a fixed load voltage is the 75% of the maximum available one $P_{max}$, with a 25% value of $\lambda_{waste}$. If the acceleration amplitude halves ($A_{max}/A_{max-0} = 0.5$), the extracted power $P_{load-V0}$ is even null with a 100% value of $\lambda_{waste}$.

This is justified by observing that, when the acceleration amplitude becomes $A_{max} = 0.5 \cdot A_{max-0}$, without any MPPT, the load generator voltage amplitude continues to be set at the value given by (18.1), that is

$$V_{load} = \frac{\Delta \cdot A_{max-0}}{2} = \Delta \cdot A_{max} \quad (23)$$

Therefore, the two generators in the mechanical stage of the equivalent electric circuit in Fig. 1(c) have the same amplitude and phase values and no current flows in the series branch ($\alpha \cdot \dot{x}(t) = 0$). What is more, in case of $A_{max} < 0.5 \cdot A_{max-0}$, a reversal of the power flow from the load voltage generator to the harvester could even happen.

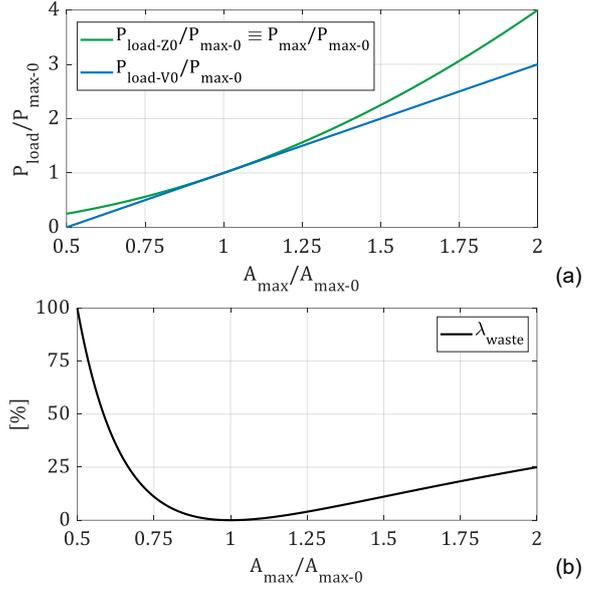

Fig. 5. (a) Trend of the powers $P_{max}$, $P_{load-Z0}$, and $P_{load-V0}$ (extracted in correspondence of $A_{max}$) normalized to the starting maximum power $P_{max-0}$ (extracted in correspondence of $A_{max-0}$) as a function of the normalized acceleration amplitude $A_{max}/A_{max-0}$. (b) Trend of $\lambda_{waste}$ as a function of $A_{max}/A_{max-0}$

The above considerations show that an electronic interface emulating a voltage generator cannot operate without an MPPT control, with the consequent increase in the control circuitry power losses compared to an electronic interface emulating a load impedance.

## 4 Proposed Load Impedance Emulation Active Interface

An AC/DC boost converter equipped with a proper control unit can be employed in order to emulate an optimal load of a piezoelectric harvester and to maximize the power extraction [13]-[19]. Differently from the electronic interfaces emulating an optimal load generator and equipped with a digital control unit implementing MPPT techniques, according to previous theoretical results, here an active interface emulating an optimal load impedance and equipped with an analog control unit is presented, as shown in Fig. 6(a).

The details of the AC/DC boost converter and of the analog control unit are shown in Fig. 6(b). The analog control unit implements a feedback loop that measures the current drawn by the electronic interface, which is the RPVEH load current $i_{load}(t)$, through the series resistance $R_m$, to make it as much as possible equal to the desired value. The control unit is made up of a conditioning stage, a hysteretic controller and a driving circuit ensuring the blanking time.

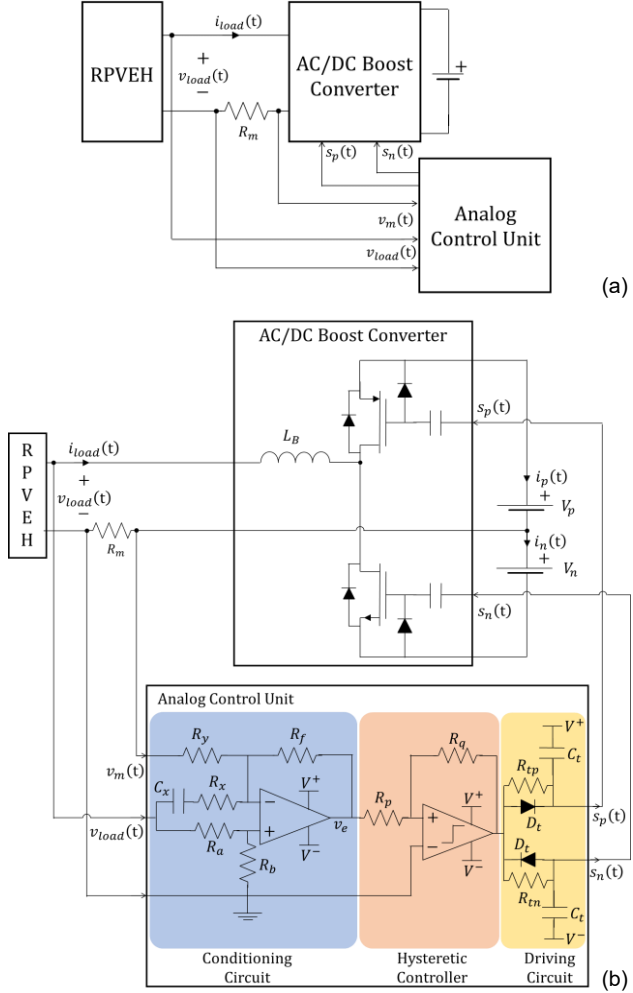

Fig. 6. (a) Schematic representation of the proposed active AC/DC interface. (b) Circuital scheme of the proposed active AC/DC interface.

The hysteretic controller is composed of a comparator and two positive feedback resistors $R_p$ and $R_q$. Since the output voltages of the comparator are the supply voltages $V^+$ and $V^- = -V^+$, the lower and higher input thresholds of the hysteretic comparator are equal to

$$V_{TH} = +\frac{R_p}{R_q}V^+; \quad V_{TL} = \frac{R_p}{R_q}V^- = -\frac{R_p}{R_q}V^+ \quad (24)$$

Thus, the hysteresis band results in

$$\Delta V_T = V_{TH} - V_{TL} = 2\frac{R_p}{R_q}V^+ \quad (25)$$

The feedback loop keeps the input signal of the hysteretic controller, i.e. $v_e$, within the hysteresis band $\Delta V_T$.

The voltage $v_e$ is the output of the conditioning circuit and can be calculated by applying the superposition principle and the virtual short-circuit condition to the op-amp

$$\bar{V}_e = -\frac{R_f}{R_y}\bar{V}_m - \frac{R_f}{\dot{Z}_x}\bar{V}_{load} + \\ +\frac{R_b}{R_a+R_b}\left(1+\frac{R_f}{R_y||\dot{Z}_x}\right)\bar{V}_{load} \quad (26)$$

with $\dot{Z}_x = R_x + 1/(j\omega C_x)$, $\bar{V}_e$, $\bar{V}_{load}$ and $\bar{V}_m$ the phasors of $v_e, v_{load}$ and $v_m$, respectively. Taking into account that $v_m = R_m \cdot i_{load}$, it results

$$\bar{V}_e = -\frac{R_f\,R_m}{R_y}\bar{I}_{load} - \frac{j\omega\,R_f C_x}{1+j\omega\,R_x C_x}\bar{V}_{load} + \\ +\frac{R_b}{R_a+R_b}\left(1+\frac{R_f}{R_y}\cdot\frac{1+j\omega\,R_x C_x+R_y C_x}{1+j\omega\,R_x C_x}\right)\bar{V}_{load} \quad (27)$$

If $\omega_x = 1/(R_x C_x)$ and $\omega_y = 1/(R_y C_x)$ are sufficiently higher than the angular frequency $\omega$ of the voltage $v_{load}$ i.e., $\omega \ll \omega_x$ and $\omega \ll \omega_y$, (27) can be simplified as

$$\bar{V}_e = -\frac{R_f\,R_m}{R_y}\bar{I}_{load} - j\omega\,R_f C_x\,\bar{V}_{load} + \\ +\frac{R_b}{R_a+R_b}\left(1+\frac{R_f}{R_y}\right)\bar{V}_{load} \quad (28)$$

The voltage $v_e$ is kept within the hysteresis band $\Delta V_T$ by the feedback loop. For a hysteresis band sufficiently small, it can be assumed $v_e \cong 0$ and thus (28) leads to

$$\frac{\bar{I}_{load}}{\bar{V}_{load}} = \frac{R_b}{R_a+R_b}\frac{R_f+R_y}{R_f R_m} - j\omega\frac{R_y C_x}{R_m} \quad (29)$$

Expression (29) shows that the controller allows the electronic interface to behave like the parallel connection of a resistance $R_e$ and of a reactance $X_e$ whose values are

$$R_e = \frac{R_a+R_b}{R_b}\frac{R_f R_m}{R_f+R_y} \quad (30.1)$$

$$X_e = \frac{R_m}{\omega\,R_y C_x} \quad (30.2)$$

Therefore, the proposed AC/DC interface is able to emulate at its input terminals the optimal RPVEH load impedance expressed by (11). Accordingly, the circuit parameters are chosen on the basis of the following relations

$$\frac{R_a+R_b}{R_b}\frac{R_f R_m}{R_f+R_y} = \frac{D}{\alpha^2} \quad (31.1)$$

$$\frac{R_y C_x}{R_m} = C_p \quad (31.2)$$

It is interesting to observe that the proposed active interface emulates at its input terminals an impedance that is the parallel of a resistance and a positive reactance obtained by means of an equivalent negative capacitance, $C_n = -C_x R_y/R_m$. The emulation of a negative capacitance in place of an inductance for obtaining a positive reactance makes the proposed interface more versatile. The value of an inductance giving the optimal reactance expressed by (13.2), i.e. $L_{opt} = 1/(\omega_{res}^2 C_p)$, would depend not only on $C_p$ but also on $\omega_{res}$. Therefore, for every mechanical configuration a different value of the inductance should be set. Differently, the value of a negative capacitance giving the optimal reactance expressed by (11.2), i.e. $C_{opt} = -C_p$, does not depend on $\omega_{res}$ but only on $C_p$ [12]. Hence, for a given RPVEH, the optimal value of the emulated negative capacitance is the same for any mechanical configuration.

Moreover, note that the proposed interface is able to emulate the optimal impedance by exploiting an analog control circuit without any MPPT control, with a beneficial effect on the losses, and, as it will be shown in the next section, also on the extracted energy.

## 5 Experimental Results

In this section the proposed load impedance emulation interface is experimentally tested and compared with active interfaces proposed in the literature.

### 5.1 Identification of the harvester parameters

The experimental tests were carried out by using a commercial RPVEH (MIDE PPA-4011) driven by mechanical vibrations produced by the shaker Sentek VT-500. A picture of the experimental setup is reported in Fig. 7(a) and a zoom showing the details of the RPVEH mounted on the shaker is shown in Fig. 7(b). The mechanical resonance frequency of the harvester mounted in this configuration was identified as the frequency for which the short-circuit current gets its maximum value and is equal to $f_{res} = 137.6\ Hz$. Moreover, the output capacitance, equal to $C_p = 405\ nF$, was measured by using the LCR meter U1733C by Keysight Technologies. Further, in order to identify the values of the parameters $\Delta$ and $\rho$ of the considered RPVEH, a load voltage generator was connected to its terminals as shown in Fig. 2(b). The average power $P_{V-load}$, extracted from the RPVEH and provided to the load voltage generator, was measured as a function of the amplitude $V_{load}$ and the phase $\Phi_{load}$ of the load voltage, in presence of an acceleration amplitude $A_{max} = 1\ g$, as reported in the surface shown in Fig. 8.

From the surface it is possible to identify the value of the optimal load voltage amplitude, $V_{opt} = 4\ V$, and the value of the maximum power, $P_{max} = 3.1\ mW$.

On the basis of such values, by exploiting (16.1) and (8), it is possible to obtain

$$\Delta = \frac{2 \cdot V_{opt}}{A_{max}} = 8\ \frac{V}{g} \quad (32.1)$$

$$\rho = \frac{(\Delta \cdot A_{max})^2}{8} \cdot \frac{\omega_{res} C_p}{P_{max}} = 0.9 \quad (32.2)$$

The obtained value of $\rho$, together with the values of $f_{res}$ and $C_p$, permits the estimation of the optimal load resistance and reactance given by (11), that is $R_{opt-TH} = \rho/(2\pi f_{res} C_p) = 2570\ \Omega$ and $X_{opt-TH} = 1/(2\pi f_{res} C_p) = 2856\ \Omega$.

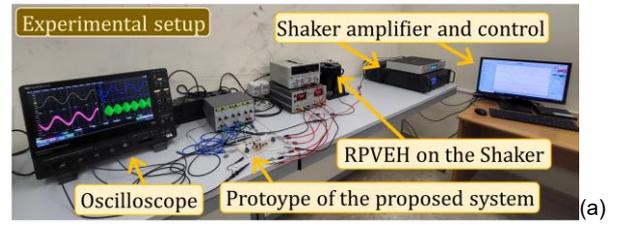
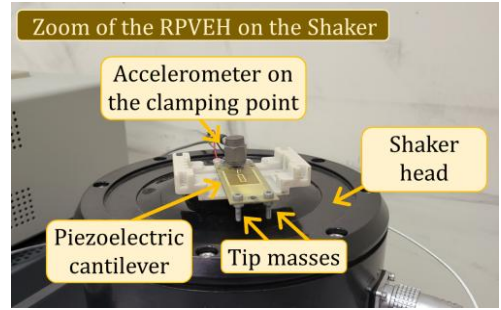
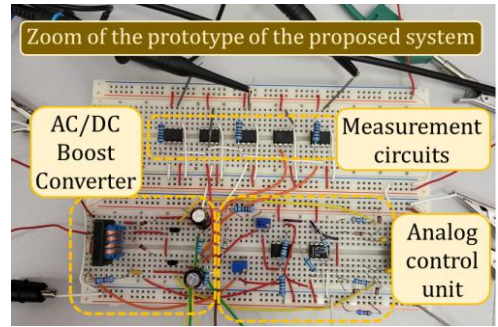

Fig. 7. (a) Image of the experimental setup. (b) Zoom of the RPVEH mounted on the shaker. (c) Zoom of the prototype of the proposed electronic interface

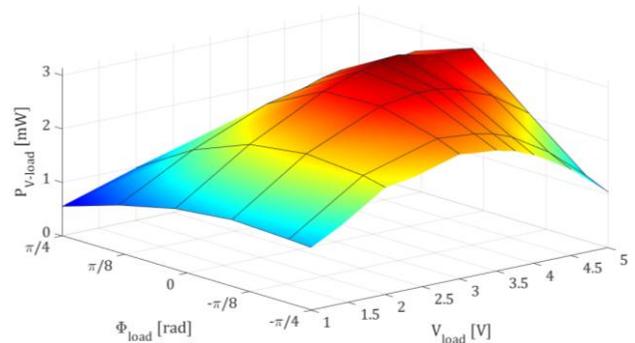

Fig. 8. Surface of the measured average extracted power as a function of $V_{load}$ and $\Phi_{load}$ with $A_{max} = 1\ g$.

The above measurements on the RPVEH connected to a load voltage generator were repeated for two other acceleration amplitudes, i.e. $0.75\ g$ and $1.25\ g$, as shown in Fig. 9. For all the three considered acceleration amplitudes, the optimal phase is nearly zero, as predicted by (16.2). The measured optimal amplitudes vary in the three cases according to (16.1) since the maximum average power is reached for $V_N = V_{load}/V_{opt}$ equal to one.

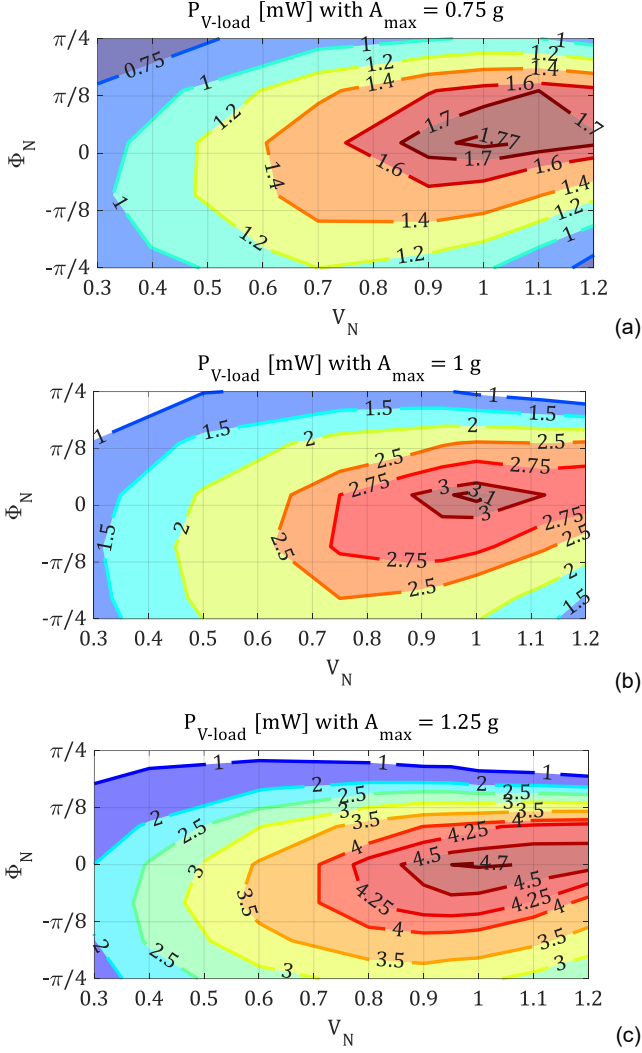

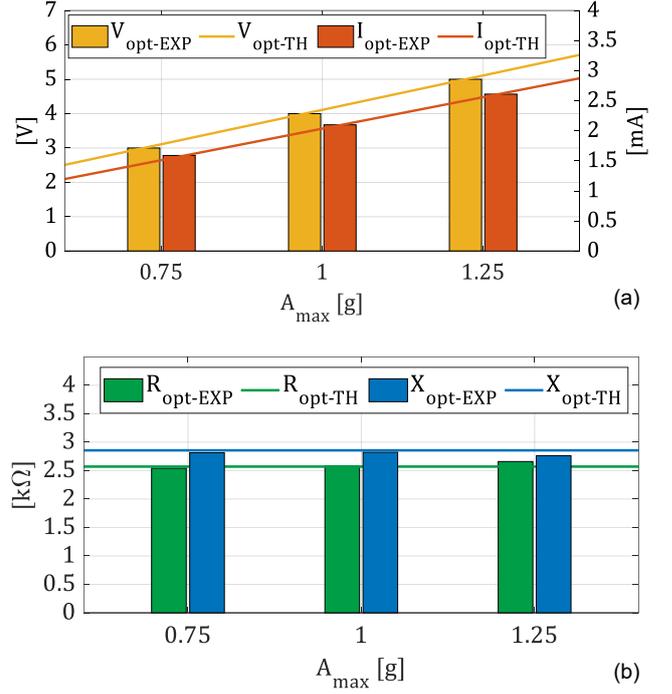

Fig. 10. Theoretical and experimental optimal operating points for the three considered accelerations. (a) Optimal voltage and current amplitudes; (b) optimal load resistance and optimal load reactance.

$$\dot{Z}_{opt} = \frac{V_{opt}}{I_{opt}}cos\theta + j\frac{V_{opt}}{I_{opt}}sin\theta \qquad (33.1)$$

$$\theta = cos^{-1}\left(\frac{2 \cdot P_{max}}{V_{opt} \cdot I_{opt}}\right) \qquad (33.2)$$

Thus, the emulated parallel load resistance $R_{opt}$ and reactance $X_{opt}$ result equal to

$$R_{opt} = \frac{Re\{\dot{Z}_{opt}\}^2 + Im\{\dot{Z}_{opt}\}^2}{Re\{\dot{Z}_{opt}\}} \qquad (34.1)$$

$$X_{opt} = \frac{Re\{\dot{Z}_{opt}\}^2 + Im\{\dot{Z}_{opt}\}^2}{Im\{\dot{Z}_{opt}\}} \qquad (34.2)$$

In Fig. 10(b), the values previously obtained from the RPVEH identification on the basis of (10), i.e. $R_{opt-TH} = 2570\ \Omega$ and $X_{opt-TH} = 2856\ \Omega$, are compared with the ones obtained on the basis of (34), as a function of $A_{max}$. The results in Fig. 10(b) confirm the theoretical prediction in Section 3, that is the values of $R_{opt}$ and $X_{opt}$ do not depend on the acceleration amplitude. Thus, also the possibility to avoid an MPPT in the case of a power electronic interface emulating the optimal load impedance is confirmed.

Fig. 9. Contour plots of the average power $P_{load}$ as a function of the load voltage phase $\Phi_{load}$ and of the normalized load voltage amplitude $V_{load}/V_{opt}$. (a) $A_{max} = 0.75\ g$; (b) $A_{max} = 1\ g$; (c) $A_{max} = 1.25\ g$.

Thus, the measurements confirm the theoretical predictions in (16) for the optimal values of the load voltage generator.

Fig. 10(a), where the measured optimal load voltage and current amplitudes are compared to the theoretical ones as a function of $A_{max}$, highlights the linear dependence of the optimal $V_{load}$ on $A_{max}$, and hence, also the necessity for an MPPT in the case of a power electronic interface emulating the load voltage generator case.

Moreover, by exploiting the measured values of $V_{opt}$, $I_{opt}$ reported in Fig. 10(a) and of $P_{max}$ reported in Fig. 9, it is possible to identify the emulated load impedance in correspondence of the optimal load voltage for each acceleration amplitude as

## 5.2 Performance of the proposed interface

A first set of experimental tests was carried out in presence of a constant acceleration amplitude to show the ability of the proposed load impedance emulation interface to allow the RPVEH to extract the maximum available power. To this aim, the harvester tested in the previous section was connected to the prototype of the electronic interface shown in Fig. 7(c). The values of the circuit parameters, shown in Table 1, leads to $\Delta V_T = 150\ mV$, $f_x = 482\ Hz$, $f_y = 19.9\ kHz$, $R_e = 2565\ \Omega$, and $C_n = -400\ nF$, according to the theoretical derivations.

Table 1
Values of the parameters of the proposed Active Interface

| Component | Value | Component | Value |
|---|---|---|---|
| $L_B$ | $100\ mH$ | $C_b$ | $100\ nF$ |
| $R_m$ | $20\ \Omega$ | $R_{tp}$ | $30\ k\Omega$ |
| $C_x$ | $1\ nF$ | $R_{tn}$ | $100\ k\Omega$ |
| $R_x$ | $330\ k\Omega$ | $C_{dt}$ | $100\ pF$ |
| $R_y$ | $8\ k\Omega$ | NMOS | ZVN4424A |
| $R_f$ | $100\ k\Omega$ | PMOS | ZVP4424A |
| $R_a$ | $275\ k\Omega$ | Diodes | 1N4148 |
| $R_b$ | $2\ k\Omega$ | OP-AMP | MCP6241 |
| $R_p$ | $150\ k\Omega$ | Comparator | LTC1440 |
| $R_q$ | $10\ M\Omega$ | $V_{DC} = V_n = V^+ = -V^-$ | $5\ V$ |

A sinusoidal vibration with three different acceleration amplitudes $A_{max}$ ($0.75\ g, 1\ g, 1.25\ g$) was applied to the RPVEH. In Figures 11(a), 12(a) and 13(a), oscilloscope screenshots are reported for the three considered cases to show the operation and the performance of the proposed interface. In the screenshots, the yellow curve is the RPVEH load voltage $v_{load}(t)$ (equal to the AC/DC converter input voltage). The blue curve is the input acceleration $a(t)$. The red curve is the RPVEH load current $i_{load}(t)$ (equal to the AC/DC converter input current). The green curve is the instantaneous power delivered to the DC load, i.e. $p_{DC}(t) = V_p \cdot i_p(t) + V_n \cdot i_n(t)$ (see Fig. 6). Let us note that the average power $P_{load}$ in Fig. 11(a), is equal to about $1.66\ mW$, nearly coincident with the maximum power extracted under the same acceleration during the harvester identification (Fig. 9(a)), that is $1.77\ mW$. For the case in Fig. 12(a), $P_{load}$ is equal to about $2.98\ mW$, nearly coincident with that in Fig. 9(b), that is $3.1\ mW$. For the case in Fig. 13(a), $P_{load}$ is equal to about $4.46\ mW$, very similar to the maximum power reported in Fig. 9(c), that is $4.7\ mW$.

Moreover, the zooms of the normalized input acceleration $a(t)$ and of the normalized RPVEH load voltage $v_{load}(t)$, reported in Fig. 11(b), 12(b), and 13(b), show that the voltage amplitudes agree with (16.1) and the voltage waveforms are nearly in phase with the accelerations, as predicted by (16.2).

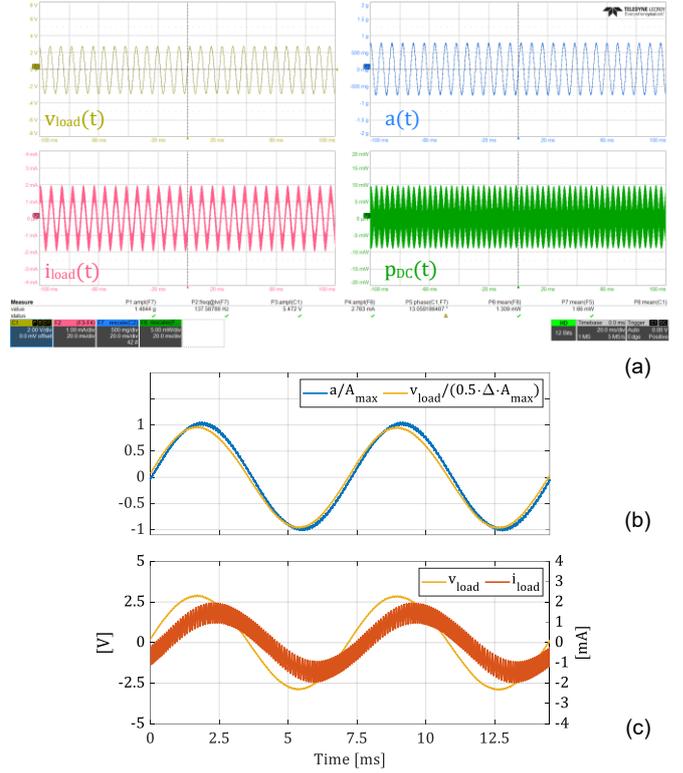

Fig. 11. Test of the proposed circuit with a constant acceleration amplitude 0.75 g

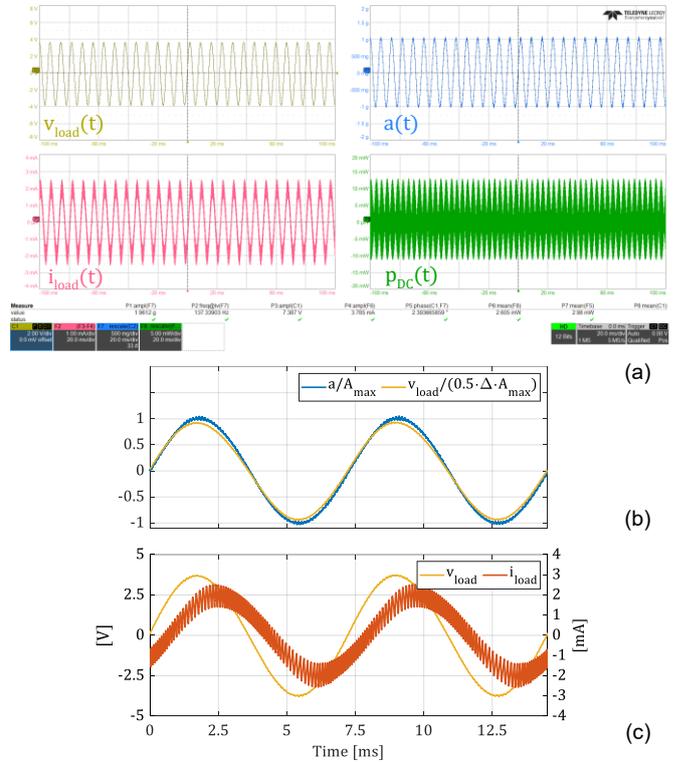

Fig. 12. Test of the proposed circuit with a constant acceleration amplitude 1 g.

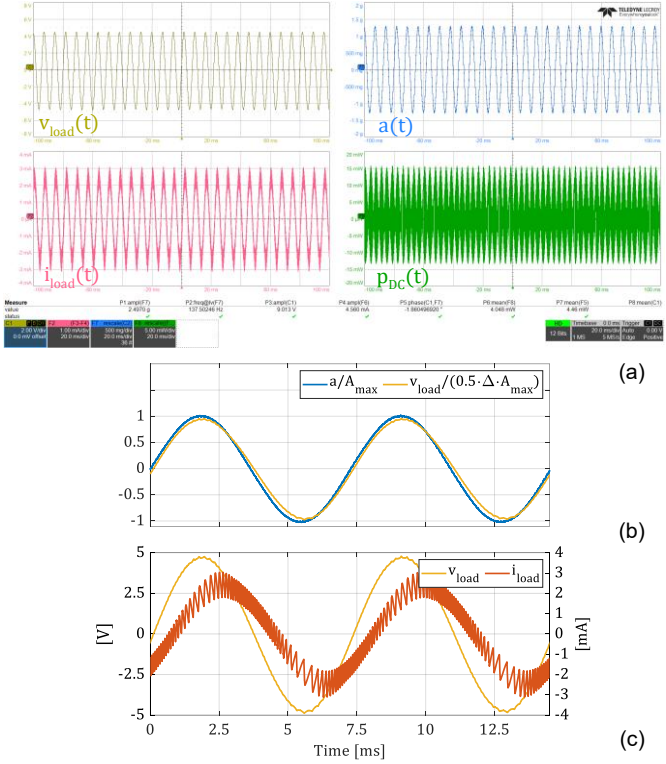

Fig. 13. Test of the proposed circuit with a constant acceleration amplitude 1.25 g.

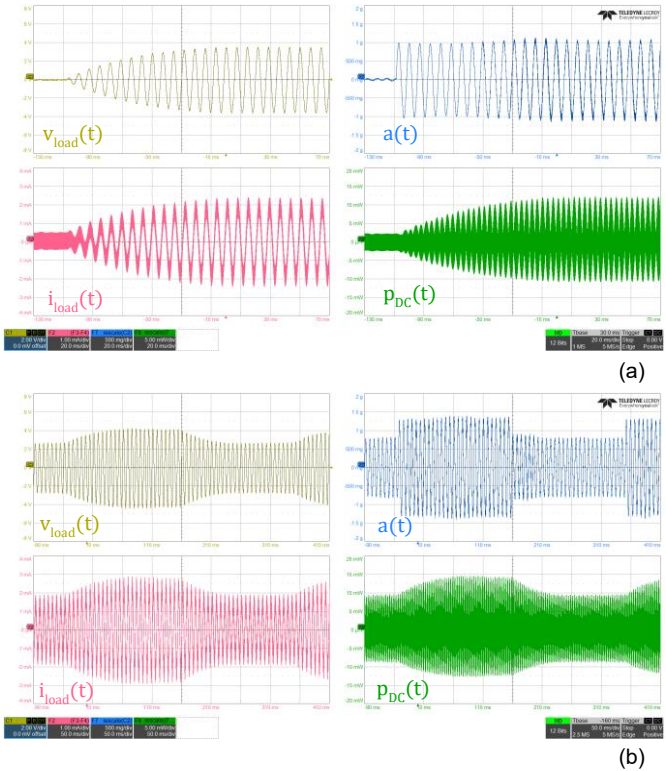

Fig. 14. Test of the proposed electronic interface under a time varying acceleration amplitude.

Therefore, for all the considered acceleration amplitudes, the proposed load impedance emulation interface allows the RPVEH to work in its optimal operating point, without any MPPT control. Finally, the zoomed load voltages and the zoomed load currents, reported in Fig. 11(b), 12(b), and 13(b), show that the proposed interface behaves like an ohmic-inductive load, with the current lagging the voltage, as it should be for a parallel of a resistance and a negative capacitance.

An additional set of experimental tests of the proposed load impedance emulation interface was carried out to test the dynamic performance of the electronic interface. Fig. 14(a) refers to the case of an amplitude variation from $0\ g$ to $1\ g$, happening at time $t = 20\ ms$. Fig. 14(b) refer to the case of a periodic variation of the acceleration amplitude $0.75\ g \rightarrow 1.25\ g \rightarrow 0.75\ g \rightarrow 1.25\ g$. The measured waveforms (with the same meaning of the colors as before) show that the proposed interface is not deceived in presence of variations of the amplitude of the vibrations, and that, after any variation, it is able to very quickly reach the new optimal operating point. This is confirmed by comparing the steady state parts of the waveforms in Fig. 14 with the corresponding waveforms in Fig. 11(a), 12(a), and 13(a).

## 5.3 Comparison with other active interfaces

Many active interfaces, based on AC/DC switching converters and able to lead to the maximum power extraction from vibration harvesters have been proposed in literature [13]-[19]. Their main characteristics are compared with those of the proposed interface in Table 2. In all the cases, the objective of the design is a best thread-off between the maximization of the tracking ability and the minimization of the complexity aimed at minimizing the losses. The main difference between the interfaces proposed in literature and the one here presented is the implementation of an MPPT algorithm, which significantly affect the harvesting performance in at least three aspects.

Firstly, the implementation of an MPPT algorithm requires the use of a microcontroller, which in turn can significantly reduce the system efficiency due to its power losses. Differently, the proposed electronic interface allows the extraction of the maximum power through an AC/DC switching converter exploiting a very light analog control, without the need for a microcontroller.

Moreover, the MPPT algorithms can be very efficient in tracking the optimal operating point of harvesters, but they are typically characterized by steady state oscillations around the optimal point, at the expense of less power extracted. Regardless of the specific characteristics of the algorithm, all the MPPT algorithms at steady state need to periodically perturb the harvester operating point and to investigate nearby points to continuously check the optimal one, even if the input vibration is stationary.

Table 2
Comparison of the proposed active interface with other active interfaces proposed in the literature

| Paper Year | Type of control | Microcontroller based | External sensors | MPPT algorithm | MPPT complexity | Stationary Performance | Dynamic performance |
|---|---|---|---|---|---|---|---|
| [13] 2024 | Load impedance matching with MPPT | Microcontroller STM32L433CC | No | P&O (Perturb and Observe) | Low (single variable MPPT) | Medium (three points oscillations) | About 3 mW in 4.5 s. Track-speed 0.67 mW/s |
| [14] 2022 | Load impedance matching | No | Auxiliary coil | No | N/A | High (no oscillations around MPP) | N/A |
| [15] 2022 | Load impedance matching with MPPT | Microcontroller TMS320F28379 | No | MPPT monitoring battery current | Low (single variable MPPT) | Medium (oscillations around the MPP) | N/A |
| [16] 2022 | Load generator adaptation | Microcontroller STM32F401RE | Hall sensors | Adaptive control loop | Low (single variable MPPT) | Medium (oscillations around the MPP) | N/A |
| [17] 2020 | Load generator adaptation with MPPT | Microcontroller PIC18F14K50 | Small piezo device | O&O (Overturn & Observe) | Medium (simplified multi-MPPT) | Low (more than three points oscillations) | About 1 mW in 25s. Track-speed 40 µW/s |
| [18] 2019 | Load impedance matching with MPPT | dSpace real-time control hardware | No | Two-variables extremum seeking | High (gradient based multi-MPPT) | Low (more than three points oscillations) | About 480 mW in 700 s. Track-speed 0.68 mW/s |
| [19] 2019 | Load generator adaptation | Only simulation results. | No | Two-variables P&O | High (many directions investigated) | Low (more than three points oscillations) | About 3 mW in 40. Track-speed 75 µW/s |
| This paper | Load impedance matching | No | No | No | N/A | High (no oscillations around MPP) | About 3 mW in 100 ms. Track-speed 30 mW/s |

This behaviour leads to oscillations of the operating point around the optimal one with a reduction in the mean extracted power. Differently, the proposed electronic interface always works exactly in the optimal point, with an increase in the mean extracted power.

Finally, the perturbative approach of the MPPT algorithms leads to limited tracking speeds under time varying acceleration, at the expense of the dynamic performance. An MPPT algorithm investigates a number $N$ of operating points before reaching the optimal operating point. The investigation of each point requires the exhaustion of the system transient following the perturbation before any power measurement and consequent decision. Thus, an MPPT algorithm requires $N$ transient intervals to reach the optimal operating point. Differently, the proposed interface allows the system to work in the optimal operating point after only one transient.

As it is shown in Fig. 14, after a variation of the input acceleration, the system is able to reach the new optimal operating point in about $100\ ms$, that is the time needed for the system to reach the new steady state condition. Any MPPT algorithm, investigating a number $N$ of operating points, would take a time that is $N \cdot 100\ ms$ to reach the new optimal operating point and hence the tracking speed would be $N$ times lower. In particular, the tracking speed of the proposed interface is about $30\ mW/s$ (with a variation of about $3\ mW$ in $100\ ms$). Due to the perturbative approach, the tracking speeds of the interfaces proposed in the literature are significantly lower, as shown in Table 2.

## 6 Conclusion

A single stage active AC/DC interface for piezoelectric energy harvesters has been presented and experimentally tested. It emulates the optimal load impedance of an RPVEH by exploiting a simple analog control without resorting to lossy microcontrollers and MMPT algorithms. The absence of perturbative approaches allows the improvement of the stationary performances since the operating point does not oscillate around the MPP in steady state conditions. Moreover, in presence of variations of the input vibration acceleration, the absence of a perturbative research of the MPP allows the enhancement of the dynamic performance. Experimental tests of the proposed interface performed in stationary and dynamic conditions confirmed the theoretical derivations.

## Funding information

This work was supported in part by European Union—Next Generation EU in the framework of PRIN 2022 under grant 20222RWCJJ, project title Hybrid Energy hArVesting systEms for multiple and irregular ambieNt sources - HEAVEN (Finanziato dall'Unione europea - Next Generation EU, Missione 4 Componente 1 CUP B53D23002170006), in part by European Union—Next Generation EU in the framework of PRIN 2022 under grant 2022Z8C472, project title Advanced Management of PEriprosthetic joint infections based on biofilm electrical featuREs - AMPERE (Finanziato dall'Unione europea - Next Generation EU, Missione 4 Componente 1 CUP B53D23020980006) and in part by European Union—Next Generation EU in the framework of PRIN 2022 PNRR under grant P202244448, project title Vibration Energy Harvesters featuring Smart Power Electronic Interfaces towards Resilient IoT - ESPERI (Finanziato dall'Unione europea - Next Generation EU, Missione 4 Componente 1 CUP B53D23023800001).

## References

[1] A. Brenes et al., 2020, "Maximum power point of piezoelectric energy harvesters: a review of optimality condition for electrical tuning" Smart Materials and Structures 29 (3), 033001.

[2] R. D'hulst, et al.: "Power Processing Circuits for Piezoelectric Vibration-Based Energy Harvesters", IEEE Trans. on Ind. Electron., vol. 57, no. 12, pp. 4170-4177, Dec. 2010.

[3] J. Dicken, P.D. Mitcheson, I. Stoianov, E.M. Yeatman: "Power-extraction circuits for piezoelectric energy harvesters in miniature and low-power applications", IEEE Trans. on Power Electron., vol. 27, no.11, pp. 4514-4528, Nov. 2012.

[4] Datasheet available online: http://www.linear.com/product/LTC3588-1.

[5] Datasheet available online: http://www.linear.com/product/LTC3331.

[6] Datasheet available online: http://www.ti.com/tool/TIDA-00690#0.

[7] Datasheet available online: https://datasheets.maximintegrated.com/en/ds/MAX17710.pdf

[8] Guyomar, D., Badel, A., Lefeuvre, E., and Richard, C. "Toward energy harvesting using active materials and conversion improvement by nonlinear processing", IEEE Trans. Ultrason. Ferroelectr. Freq. Control, vol. 52, pp. 584–595, 2005.

[9] L. Wu, X. Do, S. Lee and D. S. Ha, "A Self-Powered and Optimal SSHI Circuit Integrated With an Active Rectifier for Piezoelectric Energy Harvesting," in IEEE Transactions on Circuits and Systems I: Regular Papers, vol. 64, no. 3, pp. 537-549, March 2017.

[10] Mitcheson, P.D., Stoianov, I., and Yeatman, E.M. "Power-extraction circuits for piezoelectric energy harvesters in miniature and low-power applications", IEEE Trans. Power Electron., vol. 27, pp. 4514–4529, 2012.

[11] E. Lefeuvre et al., 2017, "Analysis of piezoelectric energy harvesting system with tunable SECE interface" Smart Mater. Struct. 26 035065.

[12] Costanzo L, Lo Schiavo A, Vitelli M. A Self-Supplied Power Optimizer for Piezoelectric Energy Harvesters Operating under Non-Sinusoidal Vibrations. Energies. 2023; 16(11): 4368. https://doi.org/10.3390/en16114368.

[13] H. Xiao, H. Peng, H. Sun, Y. Zhao, X. Liu and C. Jiang, "Automatic Impedance Matching With Dual Time-Scale P&O in Fully Self-Powered Electromagnetic Vibration Energy Harvesting," in IEEE Transactions on Power Electronics, vol. 39, no. 3, pp. 3377-3390, March 2024, doi: 10.1109/TPEL.2023.3327458.

[14] H. Xiao, H. Peng, X. Liu and H. Sun, "Fully Self-Powered Inductor-Less Electromagnetic Vibration Energy Harvesting System Using Auxiliary Coils for Hysteresis Current MPPT Control," in IEEE Transactions on Power Electronics, vol. 37, no. 11, pp. 13192-13204, Nov. 2022, doi: 10.1109/TPEL.2022.3182155.

[15] L. Wang, H. Wang, M. Fu, Z. Xie and J. Liang, "Three-Port Power Electronic Interface With Decoupled Voltage Regulation and MPPT in Electromagnetic Energy Harvesting Systems," in IEEE Trans. on Ind. Appl., vol. 58, no. 2, Mar-April 2022, doi: 10.1109/TIA.2021.3133344.

[16] L. Costanzo, A. Lo Schiavo, M. Vitelli and L. Zuo, "Optimization of AC–DC Converters for Regenerative Train Suspensions," in IEEE Transactions on Industry Applications, vol. 58, no. 2, pp. 2389-2399, March-April 2022, doi: 10.1109/TIA.2021.3136145.

[17] L. Costanzo, A. Lo Schiavo and M. Vitelli, "Active Interface for Piezoelectric Harvesters Based on Multi-Variable Maximum Power Point Tracking," in IEEE Transactions on Circuits and Systems I: Regular Papers, vol. 67, no. 7, pp. 2503-2515, July 2020, doi: 10.1109/TCSI.2020.2977495.

[18] H.K. Seyed, M. Mehrdad, A. Siamak, "A two-variable extremum seeking controller with application to self-tuned vibration energy harvesting", Smart Materials and Structures, DOI:10.1088/1361-665X/ab028b, 2019.

[19] L. Costanzo and M. Vitelli, "Two-Dimensional P&O MPPT Technique for Piezoelectric Vibration Energy Harvesters," 2019 Inter. Conf. on Clean Electrical Power (ICCEP), Otranto, Italy, 2019, pp. 227-234, doi: 10.1109/ICCEP.2019.8890146.